\documentstyle[12pt,fleqn]{article}
\def\be{\begin{equation}}
\def\ee{\end{equation}}
\def\ba{\begin{eqnarray}}
\def\ea{\end{eqnarray}}

\def\la{\mathrel{\mathpalette\fun <}}

\def\fun#1#2{\lower3.6pt\vbox{\baselineskip0pt\lineskip.9pt
        \ialign{$\mathsurround=0pt#1\hfill##\hfil$\crcr#2\crcr\sim\crcr}}}

\begin{document}
\begin{titlepage}
\null\vspace{-62pt}
\begin{flushright}UFIFT-HEP-93--2\\
May, 1993
\end{flushright}

\vspace{.2in}
\baselineskip 24pt
\centerline{\large \bf{Axion detection in the milli-eV mass range}}

\vspace{.5in}
\centerline{Pierre Sikivie, D.B. Tanner, and Yun Wang}
\vspace{.2in}
\centerline{{\it Physics Department, University of Florida,
Gainesville, FL 32611}}

\vspace{.7in}
\centerline{\bf Abstract}
\begin{quotation}

We propose an experimental scheme to search for galactic halo
axions with mass $m_a \sim 10^{-3}$eV, which is above the
range accessible with cavity techniques.
The detector consists of
a large number of parallel superconducting wires embedded in a material
transparent to microwave radiation.
The wires carry a current configuration which produces
a static, inhomogeneous magnetic field $\vec{B}_0(\vec{x})$ within the detector
volume. Axions which enter this volume may convert to photons.
We discuss the feasibility of the detector and its sensitivity.

\end{quotation}

\end{titlepage}

\baselineskip=24pt

The axion has remained a prime candidate for dark matter [\ref{a1}].
Current constraints on the
axion allow masses between $10^{-3}$eV and $10^{-7}$eV.
If the galactic halo is made up exclusively of axions, their
density in the solar neighborhood is approximately
$0.5\times 10^{-24}$gr/cm$^3$ and their velocity dispersion is
approximately $10^{-3}c$.
Galactic halo axions can be detected by stimulating their conversion to
photons in an electromagnetic cavity permeated by a strong magnetic
field [\ref{a2}]. Detectors of this type are being built
with increasing sensitivity [\ref{cav}].
However, it appears at present that these cavity detectors cannot
cover the entire mass window. In particular, their range is limited
in the direction of large axion masses by the
complexities involved in segmenting a given magnetic volume into
many small cavities. The most complex
system envisaged so far would reach $m_a \simeq 1.6\times 10^{-5}$eV
[\ref{cav}].
Much larger masses ($\sim 10^{-3}$eV) are difficult for
the cavity detector to access given
presently available technology.
In this letter, elaborating on earlier ideas [\ref{cavnew}],
we propose an alternative approach which is specifically intended
to address the possibility of larger axion masses.

The coupling of the axion to two photons is [\ref{a1}]
($\hbar=c=1$)
\be
\label{eq:coup}
{\cal{L}}_{a\gamma\gamma} = \frac{\alpha}{8\pi} \frac{a}{f_a}
\left[ \frac{N_e}{N} - \left( \frac{5}{3} +
\frac{m_d-m_u}{m_d+m_u} \right) \right]
F_{\mu\nu}\tilde{F}^{\mu\nu} \equiv
\frac{\alpha}{4\pi} \frac{a}{f_a} g_{\gamma}
F_{\mu\nu}\tilde{F}^{\mu\nu},
\ee
where $\alpha$ is the fine structure constant,
$a$ is the axion field, $f_a$ is the axion decay constant,
$m_u$ and $m_d$ are the up and down quark current masses,
and $N$ and $N_e$ are model-dependent coefficients.
In grand unified axion models, one has $N_e/N=8/3$, hence
$g_\gamma = m_u/(m_u+m_d) \simeq 0.36$.
The axion mass is given by
\be
m_a = \frac{f_{\pi} m_{\pi}}{f_a} \frac{\sqrt{m_um_d}}{m_u+m_d}
\simeq 0.6\,{\rm eV} \left( \frac{10^7{\rm GeV}}{f_a} \right).
\ee
Thus Eq.(\ref{eq:coup}) can be rewritten
\be
\label{eq:EB}
{\cal L}_{a\gamma\gamma} = -g_{\gamma} \frac{\alpha}{\pi}
\, \frac{m_a}{0.6\times 10^{16} ({\rm eV})^2} \, a \vec{E} \cdot \vec{B}.
\ee

Because of the coupling of Eq.(\ref{eq:coup}), axions will convert
to photons (and vice versa) in an externally applied magnetic field.
The cross-section for $a \rightarrow \gamma$ conversion in a region
of volume $V$ permeated by a static magnetic field
$\vec{B}_0(\vec{x})$ is [\ref{a2}]
\be
\label{eq:sigma}
\sigma = \frac{1}{16\pi^2\beta_a} \left( \frac{\alpha g_{\gamma}}{\pi f_a}
\right)^2 \int d^3 k_{\gamma} \delta(E_a-\omega) \left|
\int_V d^3x \, e^{i(\vec{k}_\gamma-\vec{k}_a)\cdot \vec{x} }
\vec{n} \times \vec{B}_0(\vec{x}) \right|^2,
\ee
where $(E_a, \vec{k}_a)= E_a(1,\vec{\beta}_a)$ is the axion 4-momentum,
and $(\omega, \vec{k}_{\gamma})= \omega(1, \vec{n})$ is the photon 4-momentum.
$E_a=\omega$ because the magnetic field is static.
The momentum transfer $\vec{q} = \vec{k}_\gamma-\vec{k}_a$,
which is necessary because the photon
is massless while the axion is massive, is provided by the
inhomogeneity of the magnetic field. Galactic halo axions
are non-relativistic ($k_a \sim 10^{-3} m_a$).
Hence, to obtain resonant conversion, the magnetic field should be made
inhomogeneous on the length scale $m_a^{-1}$.

Figure 1 shows top and side views of the detector we propose.
It consists of an array of parallel
superconducting wires embedded in a microwave-transparent
dielectric. The dielectric medium keeps the wires in place.
The dimensions of the detector are $(L_x, L_y, L_z)$.
$\hat{y}$ is the common direction of the wires. The intersections of
the wires with the $(x,z)$ plane form an array with unit cell
size $d \la m_a^{-1}$. We denote the location of a wire with
the integers $(n_z, n_x)$, where
\[
n_z \in (-N_z/2, N_z/2), \hskip 2cm N_zd=L_z;
\]
\[
n_x \in (-N_x/2, N_x/2), \hskip 2cm N_xd=L_x.
\]
Let the wires carry the following current configuration:
\be
\label{eq:I}
I(n_z,n_x)=I(n_z)=I_0 \sin(n_z d\, q).
\ee
Resonant conversion occurs when $q \sim m_a^{-1}$.
For an infinite distribution of current,
the magnetic field generated is $\vec{B}(z) = -\hat{x}\,B_0 \cos(qz)$,
where $B_0 =I_0/(qd^2)$.
We expect the magnetic field to be dominated by this term,
but since our detector has finite volume,
the field may be modified considerably by finite size effects.
We first investigate these finite size effects.

The magnetic field at point $(z, x)$ inside the detector
is dominated by its $\hat{x}$ component
\be
\label{eq:Bsum}
B_x(z, x) = \frac{I_0}{2\pi} \sum^{N_z/2}_{n_z=-N_z/2} \sin(qdn_z)
\sum^{N_x/2}_{n_x=-N_x/2} \frac{z-n_zd}{(z-n_zd)^2+
(x-n_xd)^2}.
\ee
Replacing sums with integrals, we find
\be
\label{eq:Ba}
B_x(z,x)  \simeq  -B_0 \left[ f(x) \cos(qz) -\frac{1}{2}
g(x,z) \cos\left( \frac{qL_z}{2}\right) +{\cal{O}}\left(
\frac{1}{qL} \right) \right],
\ee
for $|x| \leq L_x/2$ and $|z| \leq L_z/2$, where
\ba
\label{eq:b0}
B_0 & \equiv & \frac{I_0}{qd^2}, \nonumber\\
f(x) & \equiv & 1-e^{-qL_x/2}\, \cosh(qx),  \nonumber\\
g(x,z) & \equiv & \frac{1}{\pi} \left[ \arctan\left(
\frac{L_x/2-x}{L_z/2-z}\right) +
\arctan\left( \frac{L_x/2-x}{L_z/2+z}\right) +\nonumber \right.\\
& & \,\,\,\, \left.\arctan\left(\frac{L_x/2+x}{L_z/2-z}\right)+
\arctan\left(\frac{L_x/2+x}{L_z/2+z}\right) \right].
\ea
Note that $f(x)=1$ everywhere inside the detector volume, except
within a distance $\Delta x \sim q^{-1} \sim m_a^{-1}$
from the surface. Eq.(\ref{eq:Ba}) shows that the most important
finite size effects occur when $\cos(qL_z/2)\neq 0$.
Figures 2 and 3 show the $z$ and $x$ dependences of $B_x$
when $|\cos(qL_z/2)|=1$, the least favorable case.
Both the exact (Eq.(\ref{eq:Bsum})) and
the approximate (Eq.(\ref{eq:Ba})) expressions
for $B_x(z,x)$ are plotted.
There is excellent agreement between the two.
Of course, the exact curve displays
the kinks in $B_x$ which result from
the discreteness of the current distribution, whereas
the other curve is smooth.

Henceforth, we will make the simplifying assumption
that $B_x$ depends only on $z$.
For $m_aL_x, m_aL_y \gg 1$, Eq.(\ref{eq:sigma}) becomes
\be
\sigma = \sigma_0  \sum_{n_{z}=\pm 1}
 \left| \frac{1}{L_z} \int_{-L_z/2}^{L_z/2}
dz \frac{B_x(z)}{B_0} \, e^{iE_a(n_{z}-\beta_{az}) z} \right|^2,
\ee
where we have dropped terms of ${\cal{O}}(\beta_a^2)$,
and $\sigma_0$ is defined by
\be
\sigma_0 = \frac{1}{4\beta_a}
\left( \frac{\alpha g_\gamma}
{\pi f_a} \right)^2 L_x L_y L_z^2 \, B_0^2.
\ee
For $B_x(z) = -B_0 \cos(qz)$,
\be
\label{eq:res}
\sigma =\frac{\sigma_0}{L_z^2}
\left\{ \left( \frac{\sin \left\{ \frac{L_z}{2}[E_a(1+\beta_{az})-q]
\right\}} {E_a(1+\beta_{az})-q} \right)^2 +
\left( \frac{\sin\left\{\frac{L_z}{2}[E_a(1-\beta_{az})-q]\right\}}
{E_a(1-\beta_{az})-q} \right)^2 \right\},
\ee
where only the resonance terms are kept. The purpose of the detector
is to search for the axion signal resonance by tuning the
wave number
$q$ of the current configuration. Eq.(\ref{eq:res}) shows
that the bandwidth of the detector is $\Delta k^d_{z} \simeq 2/L_z$.
On the other hand, the width of the axion signal
in the same variable is $\Delta k^a_{z} \simeq 2\times
10^{-3}m_a$. Provided that the axion signal falls entirely within
the bandwidth of the detector $(|q-m_a \pm 10^{-3}m_a | <1/L_z)$,
the power into the detector from $a\rightarrow \gamma$ conversion on
resonance ($q\simeq m_a$) is
\ba
P & = & \sigma \beta_a \rho_a = \frac{1}{8} \left( \frac{\alpha g_\gamma}
{\pi f_a} \right)^2 V L_z B_0^2 \rho_a \nonumber\\
& = & 2 \times 10^{-21} {\rm Watt} \left(\frac{VL_z}{(
{\rm meters})^4}\right)\left(\frac{B_0}{8 \,
{\rm Tesla}} \right)^2 \left(\frac{m_a}{10^{-3}{\rm eV}}
\right)^2  \nonumber\\& & \,\,\,
\left( \frac{\rho_a}{0.5\times 10^{-24} {\rm gr}/{\rm cm}^3} \right)
\left( \frac{g_{\gamma}}{0.36} \right)^2,
\ea
where $V = L_xL_yL_z$.
This power must be collected and brought to the front end of a
microwave receiver. Let $\zeta$ be the efficiency with which
this can be done and let $T$ be the total
(physical plus electronic) noise temperature of the receiver. The signal
to noise ratio over the frequency bandwidth $\Delta f_s = 10^{-6}m_a/2\pi$
of the axion signal is
\be
\frac{s}{n} = \frac{\zeta P}{T} \sqrt{\frac{t}{\Delta f_s}},
\ee
where $t$ is the measurement integration time.
The search can be carried out over the whole frequency bandwidth of
the detector ($\Delta f_d \simeq 2/(2\pi L_z)$) simultaneously. Thus
the search rate for a given signal to noise ratio is
\ba
\frac{\Delta m_a}{t} & = & 2\pi \frac{\Delta f_d}{t} = 2\pi
\frac{\Delta f_s}{t} \frac{\Delta f_d}
{\Delta f_s} = 2\pi \left( \frac{n}{s} \right)^2 \left(
\frac{\zeta P}{T}\right)^2 \frac{2Q_a}{m_a L_z} \nonumber\\
&=& \frac{6\times 10^{-5} {\rm eV}}{\rm year} \left( \frac{V^2 L_z}
{({\rm meter})^7} \right) \left( \frac{B_0}{8 \,{\rm Tesla}}\right)^4
\left( \frac{\rho_a}{0.5\times 10^{-24} {\rm gr}/{\rm cm}^3} \right)^2 \times
\nonumber\\
& & \,\,\, \left( \frac{\zeta}{0.3} \right)^2 \left(\frac{10{\rm K}}{T}
\right)^2 \left(\frac{m_a}{10^{-3}{\rm eV}}\right)^3 \left(\frac{4n}{s}
\right)^2 \left(\frac{g_\gamma}{0.36}\right)^4,
\label{eq:srate}
\ea
where $Q_a=10^6$ is the ``quality factor'' of the axion
signal, i.e., the ratio of energy to energy spread
of galactic halo axions.

Assuming that $T \simeq 10$K is realistic, Eq.(\ref{eq:srate})
tells us that a detector of linear dimensions of order a couple
of meters is required to search at a reasonable rate
near $m_a=10^{-3}$eV.
If $L_x=L_y=L_z=2$m and $d=0.2$mm, there
are $10^8$ wires arranged in $10^4$ planes of $10^4$ wires each.
To extend the search to smaller axion masses keeping a reasonably
large search rate, successively larger ($L \sim m_a^{-3/7}$) but
less fine grained ($d\sim m_a^{-1}$) detectors are needed.
For a search near $m_a=10^{-5}$eV, assuming again $T\simeq 10$K,
the detector needs to be of order $10$m in linear dimension
but requires only $10^5$ wires in a $d=2$cm matrix.

We now turn to some practical concerns.
The current configuration $I_y(z)$
must be chosen such that its wave number $q$ can be
easily tuned.
Since $I_y(z)$ is periodic,
the number of different-value currents  needed
is much smaller than the total number of wires in $\hat{z}$
direction. If we replace the harmonic current
function $I_y(n_z)$ with a function having the same repeat length along
$z$ but with integer  ratios of the
currents, the number of current sources
can be minimized by employing current dividers.
Figure 4 shows the magnetic field produced by the square wave,
triangular wave, and sine wave current configurations as
functions of $z$, given the {\it same} maximum current strength.
Our numerical computation for
the conversion cross section of axions into photons
as a function of $m_a$ shows
the expected resonance peak ($m_a=q$) prominent for
all three current configurations.

The radius $b$ of the wires must be chosen
neither too large nor too small.
If $b$ is too small,
the magnetic field at the surface of the wires
will exceed the critical field strength $B_{c}$
for the breaking down of superconductivity
even when the maximum {\it average} field $B_0$ is
still far below $B_c$.
The contribution $B_b = I/(2\pi b)$ on the surface of a wire,
which is due to the current $I$ in the wire itself, must be less
than $B_c$ for all $I$ up to $I_0=qd^2B_0$.
The resulting requirement on the wire radius is:
\be
b  > b_{min} = 9.5\,\mu{\rm m} \, (qd)\left( \frac{d}{0.2{\rm mm}} \right)
\left(\frac{B_0/B_c}{0.3} \right).
\ee
If the wire radius is sufficiently large compared with $b_{min}$,
the magnetic field at the surface of a wire is dominated by
the contribution from other wires (i.e., $B_0$),
and $B_c$ constrains $B_0$ directly.

If $b$ is too large, the signal from $a\rightarrow \gamma$
conversion is lost to photon scattering by the wires.
The photons emitted by axion decay all have momenta very nearly
parallel to the $z$ axis and polarization
perpendicular to the superconducting wires
($\vec{E} \perp \hat{y}$, see Eq.(\ref{eq:EB})).
The total scattering cross section per unit length
of wire for such photons
is [\ref{scat}]
\[
\sigma_T \simeq \frac{3\pi^2}{4} b (qb)^3.
\]
where $q$ is the wavenumber of the photons.
For $b=10\,\mu$m, $d=0.2mm$, and $L=2$m,
the loss of the signal power
at the very worst (when $qd=1$) is less than 20\%.
The polarization with $\vec E
\parallel \hat{y}$ is strongly reflected by the wires.
If the typical deviation from parallelism
in the orientation of individual wires in
the matrix is $\Delta\theta$, the loss of signal power
is on the order of $(\Delta\theta)^2$. This loss is
less than that due to
the scattering of photons by the wires if $\Delta\theta < 0.5\times 10^{-2}$,
which is easy to satisfy.

For $d=0.2\,$mm and $qd = 1$, $B_0 =  10\,$Tesla implies
$I_0 = 1591\,$Ampere. For $L = 2\,$meters,
the maximum force on a wire would be $\sim 3\times 10^4\,$N;
assuming $b=10\,\mu$m,
this corresponds to a pressure of 0.25$\,$GPa, safely below
the strength limit of a number of casting resins.
The requirements are similar for other values of q and d.

Finally, plane-polarized, well-collimated
photons will emerge from the ends of the detector in the $\pm \hat{z}$
direction. These can be focussed with a parabolic mirror or channeled
into a waveguide and brought to the input of a low-noise microwave
receiver. The signal consists of noise (both emitted by the detector and
generated by the reciever front end electronics) along with excess power
resulting from axion conversion. As in the case of cavity detectors, the
signal will be spectrum analyzed to search for the axion resonance as the
detector's operating frequency is swept. The electronics and signal
analysis techniques would be similar to that of the cavity detectors.

In summary, we have described a detector for dark-matter axions that
could be used to search the part of the allowed range of axion masses
that is unreachable in the foreseeable future by cavity detectors.


This work was supported in part by the Department of Energy under grant
\#DE-FG05-86ER40272 at the University of Florida.

\newpage
\frenchspacing
\parindent=0pt

\centerline{{\bf References}}

\begin{enumerate}

\item\label{a1} Recent reviews include: M. Turner, {\it Phys. Rep.}
{\bf 197} (1990) 67; P. Sikivie, ``Dark Matter Axions'',
U. of Florida preprint UFIFT-HEP-92-25 (1992).

\item\label{a2} P. Sikivie, {\it Phys. Rev. Lett.} {\bf 51} (1983) 1415,
{\it Phys. Rev.} {\bf D32} (1985) 2988; L. Krauss, J. Moody, F. Wilczek
and D. Morris, {\it Phys. Rev. Lett.} {\bf 55} (1985) 1797.

\item\label{cav} S. DePanfilis et al., {\it Phys. Rev.} {\bf D40}
(1989) 3153; C. Hagmann et al., {\it Phys. Rev.} {\bf D42} (1990) 1297;
K. Van Bibber et al., ``A proposed search for dark matter axions
in the 0.6--16 $\mu$eV range'', LLNL preprint UCRL-JC-106876 (1991).

\item\label{cavnew} P. Sikivie in Ref.[\ref{a2}];
P. Sikivie, N. Sullivan and D. Tanner, unpublished;
P. Sikivie in Ref.[\ref{a1}].

\item\label{scat} J.J. Bowman, T.B.A. Senior, and P.L.E. Uslenghi,
{\it Electromagnetic And Acoustic Scattering By Simple Shapes},
Hemisphere Publishing Corporation, New York (1987).
\end{enumerate}

\newpage
\nonfrenchspacing
\parindent=20pt

\centerline{{\bf Figure Captions}}

\vspace{0.2in}

Fig.1. Top and side views of the detector, showing the arrangement of wires.

\vskip .2in
Fig.2. $B_x(z,x)$ versus $z$ for $|x|=L_x/12$,
$I =I_0 \sin(n_z d q)$.
The spiked thin line is numerical and exact, while the smooth bold line
is the analytical result of Eq.(\ref{eq:Ba}).
$qd = \pi/20$, $L_x=L_z=600\,d$.

\vskip .2in

Fig.3. $B_x(z,x)$ versus $x$ for $|z|= L_z/6$,
$I =I_0 \sin(n_z d q)$. The spiked thin line
is numerical and exact, while the smooth bold line is the analytical
result of Eq.(\ref{eq:Ba}).
$qd = \pi/20$, $L_x=L_z=600\,d$.

\vskip .2in

Fig.4. $B_x(z)$ versus $z$ produced by the sine (middle bold line),
square (outer thin line) and triangular
(inner thin line) wave current configurations.
$qd = \pi/20$, $L_x=L_z=500\,d$.

\end{document}